\documentclass[10pt, onecolumn]{IEEEtran}
\usepackage{cite}
\usepackage{color}
\usepackage{comment}
\usepackage{graphicx}
\usepackage{amsmath}
\usepackage[mathscr]{eucal}
\usepackage{amssymb}
\usepackage{makeidx}
\usepackage{multicol}
\newcommand{\beq}[1]{\begin{equation} \label{#1} }
\newcommand{\beaq}[1]{\begin{eqnarray} \label{#1} }
\newcommand{\be}{\begin{equation}}
\newcommand{\bea}{\begin{eqnarray}}
\newcommand{\ee}{\end{equation}}
\newcommand{\eea}{\end{eqnarray}}
\newcommand{\bfi}{\begin{figure}}
\newcommand{\efi}{\end{figure}}

\newcommand{\f}[2]{\frac{#1}{#2}}

\newcommand{\eps}{\epsilon}

\newcommand{\D}{{\rm d}}

\newcommand{\bib}{\bibitem}
\begin{document}
\title{\Large\bf Reply to ``Comment on `Absence of a consistent classical equation of motion for a mass-renormalized point charge' '' (arXiv:2511.02865v1, 3 Nov 2025)}
\author{Arthur D. Yaghjian\\\textit{\normalsize Electromagnetics Research, Concord, MA  01742,  USA} ({\small a.yaghjian@comcast.net}) 
}
\maketitle
\thispagestyle{empty}
\begin{abstract}
\textbf{By means of a brief review of the derivation of the causal modified Lorentz-Abraham-Dirac classical equation of motion from the renormalization of the mass in the modified equation of motion of an extended charged sphere as its radius approaches zero, it is shown that Zin and Pylak's objection that the jumps in velocity allowed across transition intervals near nonanalytic points in time of the externally applied force produce delta functions in the radiated fields is incorrect.}
\end{abstract} 
\begin{IEEEkeywords}
\textbf{Causality, charged sphere,  classical equation of motion, momentum-energy conservation.}
\end{IEEEkeywords}
\unboldmath
\section{Introduction}
In the paper \cite{Absence} and book  \cite{Yaghjian3rd}, the equation of motion of the classical model of a relativistically rigid surface-charged spherical insulator of radius $a$ and total charge $e$ is derived from Maxwell's equations, the relativistic generalization of Newton's second law of motion, and Einstein's mass-energy relation; also see the recent synopsis of the work in \cite{arxiv}.  The derivation takes into account that the Lorentz power-series for the electromagnetic self-force is not valid for the time durations $\varDelta t_a \approx 2a/c$ just after nonanalytic points in time of the external force on the charged sphere, such as when the external force is first applied and terminated (with $c$ the free-space speed of light).  These transition time intervals are approximately equal to the time it takes light to traverse the diameter of the rest-frame sphere .  Although the fields and self-force cannot be evaluated in detail during these transition time intervals because the precise behavior of the velocity, radiation, and inertial mass are unknown during the transition intervals, there are transition forces during (and only during) the transition intervals that maintain causality of the resulting equation of motion, even as the radius $a$ of the charged sphere approaches zero and the mass $m$ is renormalized to a finite value.
\par
In their recent comment on this work \cite{Z&P}, Zin and Pylak claim that the jumps in velocity across the transition intervals produced by jumps in the externally applied force in the mass-renormalized equation of motion of the point charge ($a\to 0$) \cite{Absence},  \cite{Yaghjian3rd} modified by the transition forces imply from Maxwell's equations an unphysical infinite radiated energy.  Although this modified mass-renormalized Lorentz-Abraham-Dirac (LAD) equation of motion predicts a finite radiated energy, and the objection of Zin and Pylak was addressed in \cite[footnote 8, ch. 8 ]{Yaghjian3rd} (footnote 7 of the 2nd edition), it will be explained here in more detail that the textbook formula for the radiated energy of a point charge is not applicable to an extended charge as $a\to 0$ and the mass is renormalized.
\par
First, some other mistaken statements in the comments of Zin and Pylak will be mentioned.  They say that ``These [transition] intervals occur when the external force is nonzero only on some part of the charged sphere.''  This is not quite correct.   As explained in \cite[see, for example, p. 2]{Yaghjian3rd}, the entire charged sphere is assumed to be held fixed in an external electromagnetic field when it is released instantaneously at a time $t = t_1$.  The external force is assumed to be an analytic function of time for $t>t_1$ until it is terminated instantaneously across the entire sphere at $t=t_2$.  The two transition intervals have durations $\varDelta t_a \approx 2a/c$ just after $t_1$ and $t_2$ (in the instantaneous rest frames at $t_1$ and $t_2$).
\par
Also, equation (1) of Zin and Pylak, taken from equation (8.83) of the second edition of my book, applies only to evaluating the total energy radiated during both transition intervals, not the energy radiated during each transition interval.  Because of its limited application, this equation was omitted in the third edition of the book \cite{Yaghjian3rd}.
\section{Response to Primary Objection}
Zin and Pylak claim that the energy radiated $W_{\mathrm{TI},n}$ during the $n$th transition interval obtained from the modified LAD rectilinear equation of motion for the mass-renormalized charge as $a\to 0$, namely \cite[eq.(8.84)]{Yaghjian3rd}
\beq{825-1}
\frac{W_{\mathrm{TI},n}}{mc^2} =\frac{1}{mc^2} \int\limits_{t_n}^{t_n^+ = t_n+\varDelta t_{a\to 0}}\left (\frac{e^2}{6\pi\eps_0c^3}\gamma^6 \dot{u}^2 - f_{an}u\right) \D t =\tau_e[\gamma(t_n^+)\dot{\gamma}(t_n^+)- \gamma(t_n)\dot{\gamma}(t_n)] -[\gamma(t_n^+)-\gamma(t_n)]
\ee
where $u(t)$ is the center velocity of the sphere witn  $\gamma = (1-u^2/c^2)^{-\frac{1}{2}}$, $\tau_e = e^2/(6\pi\eps_0 mc^3)$, and $\eps_0$ is the free-space permittivity, is not valid because the integral of just the $\dot{u}^2$ term (without the transition-force term $f_{an}u$) is the textbook point-charge Maxwellian radiated energy.  Moreover, this textbook point-charge radiated energy is infinite for transition intervals with abrupt jumps in velocity (giving delta functions in acceleration).  As will be explained in detail, the ad hoc (not derived from fundamental equations) renormalization, which allows these jumps in velocity, also violates Maxwell's equations during the transition intervals but not outside the transition intervals, so that the modified LAD equation of motion gives the correct finite radiated energy in the last equality of (\ref{825-1}) from the integral in (\ref{825-1}).  Without renormalization, the mass of the charged sphere approaches an infinite value as $a\to 0$ and both the jump in velocity and radiated energy across a transition interval approach  zero.
\par
It is not surprising that the energy radiated during a transition interval in (\ref{825-1}) involves $ f_{an}u$ because \cite[eq. (8.41)]{Yaghjian3rd} shows that $f_{an}$ depends on both the external force $F_{\rm ext}$ and the acceleration $\dot{u}$ in the transition interval.  Moreover, in order to keep the same $\dot{u}^2$ term on the right-hand side of the modified LAD rectilinear power equation of motion \cite[eq. (8.80b)]{Yaghjian3rd}
\beq{6}
\frac{[F_\mathrm{ext}(t)+f_{a}(t)]u}{mc^2}  = \f{\D \gamma}{\D t} -\tau_e
 \left[ \f{\D }{\D t}\left(\gamma\frac{\D \gamma}{\D t}\right)-  \frac{\gamma^6}{c^2} \dot{u}^2 \right],\;\;\;f_a(t) = \sum_{n=1}^N f_{an}(t)
\ee
during a transition interval where the power series used to derive the right-hand side of (\ref{6}) is not valid, there has to be an effective transition force $f_{an}(t)$  that does work on the charged sphere during the transition interval and subtracts a contribution from the $\dot{u}^2$ power radiated  during the transition interval.
\par
If the mass is not renormalized as $a\to0$ so that $m = e^2/(8\pi\eps_0ac^2) \to\infty$ as $1/a$ and $\tau_e = 4a/(3c)\to 0$ as $a$, the velocity jump $\varDelta u_n$ and thus the $\dot{u}^2$-term energy radiated across a transition interval for a finite external force as determined by the Maxwellian fields from the surface charge on the extended charged sphere can be shown to approach zero as $a\to 0$.    Similarly, the energy radiated by the $f_{an}u$ term approaches zero as $a\to 0$.  However, if the mass is renormalized to a finite value $m$, the modified LAD equation of motion requires, for a jump in the external force, a nonzero jump in velocity $\varDelta u_n$ across a limitingly short transition interval in order to avoid an unphysical negative radiated energy.  A jump in velocity means that the acceleration will contain a delta function $\varDelta u_n\delta(t)$.  A naive application of the textbook point-charge Li\'enard-Wiechert potentials \cite[sec. 2.1.8]{Yaghjian3rd} then predicts a delta function in the radiated fields proportional to $\dot{u}(t)= \varDelta u_n\delta(t)$ and  infinite radiated energy [since $e^2\varDelta u_n^2\int\delta^2(t) dt = \infty$].   
\par
However, if the far fields \cite[eqs. (5.34)-(5.35)] {Yaghjian3rd} are evaluated for an extended charge of nonzero radius $a$ with  $\dot{u}(t) =\varDelta u_n \delta(t)$ of infinitesimal duration in time, the far fields extend over a time duration equal to $2a/c$ with a finite magnitude proportional to $1/(2a/c)$ so that the total energy radiated during the transition interval is proportional to $1/(2a/c)$ and not $\int\delta^2(t) dt = \infty$; this result was first obtained by Paul Hertz (Abraham's student) \cite{Hertz} and Abraham \cite[sec. 25]{Abraham} as a possible explanation of X-ray production.  
\par
Therefore, for a nonzero value of the radius $a$, neither the fields nor the energy radiated  during a transition interval are infinite even though the jump in velocity of the entire charged sphere is assumed instantaneous.  Nonetheless, as the radius $a\to 0$ and the mass is renormalized to a finite value, the energy radiated by a jump in velocity approaches infinity as $1/(2a/c)$.  The reason that this $a\to 0$ divergent radiated velocity-jump energy based on Maxwell's equations applied during the transition intervals conflicts with the finite radiated energy predicted by (\ref{825-1}) obtained from the equation of motion is the renormalization of the mass to a finite value $m$.   
\par
Renormalization as $a\to 0$ is an ad hoc unphysical alteration of the classical equation of motion that implicitly changes the $1/r^2$ variation of the Maxwellian near electric field of the limitingly small charge so that an unbounded energy of formation (electrostatic mass) is no longer produced.    Such a change to the Maxwellian near fields during a transition interval cannot occur without a change in the radiated far fields as well and thus Maxwell's equations cannot be applied to find the radiated fields during the transition intervals if $a\to 0$ and the mass is renormalized.   By changing the ratio of the coefficients of the radiation-reaction and Newtonian-acceleration forces in the equation of motion,  mass renormalization as $a\to 0$ allows jumps in velocity across transition intervals that prevent Maxwell's equations from predicting the detailed electromagnetic fields radiated during the limitingly short transition intervals.  Fortunately, one can determine the total radiated energy and momentum during the transition intervals from (\ref{825-1}) and the corresponding equation for momentum using the values of the velocity and acceleration outside the transition intervals -- a determination that does not require a detailed knowledge of how renormalization effectively changes the fields produced during the transition intervals.
\par
In summary, it is shown in \cite{Absence} and  \cite{Yaghjian3rd} that a causal classical equation of motion that satisfies Lorentz covariance and momentum-energy conservation can be derived for the charged spherical insulator from a careful rigorous application of Maxwell's equations, the relativistic generalization of Newton's second law of motion, and Einstein's mass-energy relation, provided mass renormalization is not introduced.  Even with renormalization of the mass of the charged sphere as $a\to 0$, the derived equation of motion remains valid (under the condition of relativistic (Born) rigidity and avoidance of extraordinarily large jumps in externally applied forces \cite{Absence,Yaghjian3rd}) if one realizes that renormalization precludes the use of Maxwell's equations to directly find the radiated energy and momentum during the infinitesimally short transition intervals.  Fortunately, one can still rely on the integrations of the equation of motion over the transition intervals to indirectly, but straightforwardly, obtain the energy and momentum radiated during the transition intervals.  Unfortunately, neither a classical nor quantum equation of motion exists for an actual finite-mass point charge such as the electron that avoids ad hoc mass renormalization.  Such an equation of motion would require the holy-grail unification of electrodynamic and inertial/gravitational forces that, as Dirac wrote \cite{Dirac1938}, ``brings one up against the problem of the structure of the electron, which has not yet received any satisfactory solution.''  The instantaneous jump in velocity, produced by an instantaneous jump in applied external force, of a classical charged sphere with renormalized mass as the radius of the sphere approaches zero radiates a finite energy.   However, it would be impossible to determine if an instantaneous jump in external force applied to an actual electron produced an instantaneous jump in velocity because such a rapid jump in velocity would be masked by quantum effects.
\par
Since Zin and Pylak's primary objection, namely that the classical mass-renormalized point charge (sphere radius $a\to 0$) with a jump in velocity, as given in  \cite{Absence, Yaghjian3rd}, radiates delta-function fields, is incorrect, their secondary objection based on the interaction of delta-function fields between two charged particles is also incorrect.
\end{document}